# Subsurface Depths Structure Maps Reconstruction with Generative Adversarial Networks

Dmitry Ivlev, *Petroleum Overseas ME JSC,*

dm.ivlev@gmail.com

## Abstract

This paper described a method for reconstruction of detailed-resolution depth structure maps, usually obtained after the 3D seismic surveys, using the data from 2D seismic depth maps. The method uses two algorithms based on the generative-adversarial neural network architecture. The first algorithm StyleGAN2-ADA accumulates in the hidden space of the neural network the semantic images of mountainous terrain forms first, and then with help of transfer learning, in the ideal case – the structure geometry of stratigraphic horizons. The second algorithm, the Pixel2Style2Pixel encoder, using the semantic level of generalization of the first algorithm, learns to reconstruct the original high-resolution images from their degraded copies (super-resolution technology).

There was demonstrated a methodological approach to transferring knowledge on the structural forms of stratigraphic horizon boundaries from the well-studied areas to the underexplored ones. Using the multimodal synthesis of Pixel2Style2Pixel encoder, it is proposed to create a probabilistic depth space, where each point of the project area is represented by the density of probabilistic depth distribution of equally probable reconstructed geological forms of structural images.

Assessment of the reconstruction quality was carried out for two blocks. Using this method, credible detailed depth reconstructions comparable with the quality of 3D seismic maps have been obtained from 2D seismic maps.

**Keywords:** machine learning, structural map reconstruction, generative-adversarial networks, GAN, StyleGAN2-ADA, Pixel2Style2Pixel, transfer learning, 2D seismic, 3D seismic, relief, super-resolution, multimodal synthesis, probabilistic space of plausible reconstructions.

## Introduction

Subsurface depth structure mapping is a key component of geological exploration and searching for subsurface reservoirs. Geostatistical approaches to subsurface mapping are strongly dependent on the local input data. The best quality of structural imaging is currently achieved by 3D seismic data acquisition in combination with drilling and testing of wells.  However, the most part of the prospective area is covered only by widely spaced 2D seismic lines. In view of insufficient data availability, geostatistical methods have limitations in terms of using the data from the adjacent and well-studied areas. Prior to the widespread penetration of electronic data processing into the industry, lack of information was compensated by the gut feeling of the geologist, who reconstructed the area based on his extensive experience in mapping the areas with better exploration maturity. This paper describes the approach to creation of such "gut feeling" on the basis of previous experience by subsurface depths structure maps reconstruction in the conditions of data shortage using the models of generative-adversarial networks (GAN) trained on high-resolution spatial data. In a sense, the approach presented in this paper is the return to tradition at a new technological level.

# Method

A super-resolution method - a process of image recovery from low-resolution images to high-resolution images - was used in this work. The distinctive feature of this method is that a low-resolution image contains no additional information; this information is reconstructed into a high-resolution image with help of a neural network capable of generalizing at a very high semantic level.

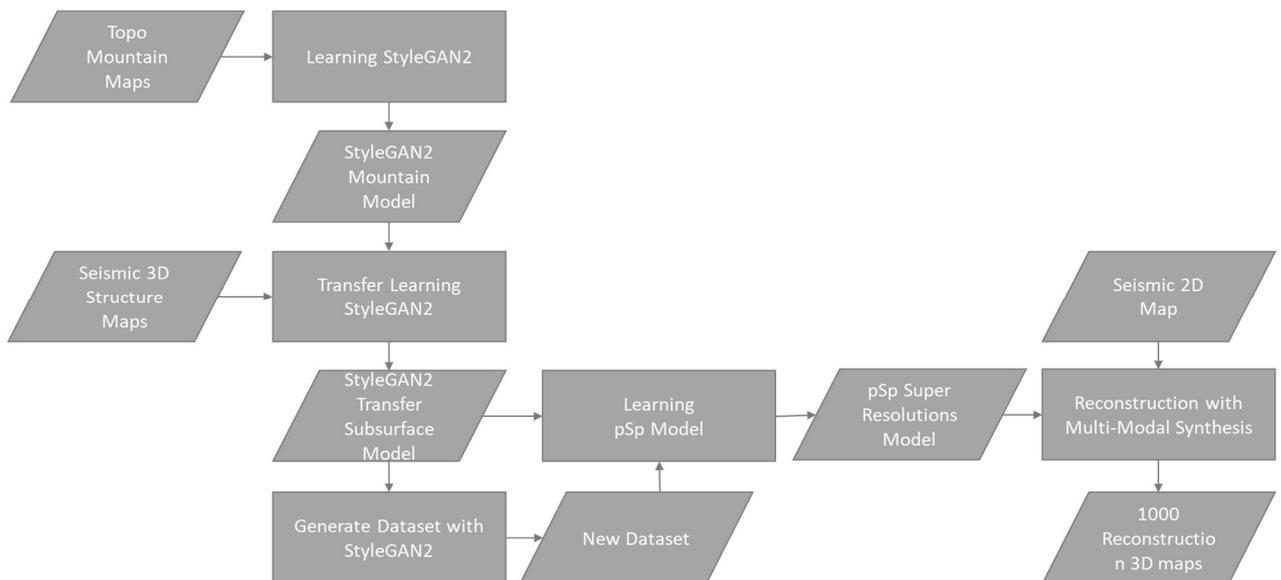

Fig. 1. Algorithm of reconstruction method implementation.

The choice of this method for reconstruction is based on the naive assumption that 2D structure maps are to some extent the 3D maps with poor sampling that can be used to reconstruct the high-resolution variants of the project area based on accumulated representations of the neural network on the possible shapes of the geological structures. Thus, one neural network forms a semantic "database" of topographic forms, and the task is for another neural network-encoder to select such a high-resolution form that would coincide with the original low-resolution image when the image sampling (degradation). From the very beginning this task is classified as ill-defined, as it does not have a single solution. However, this can be useful because, depending on the quality of input data, it allows to form a probabilistic space of possible forms from a combination of real-life forms which were used for training. The reconstruction method was implemented using two algorithms of generative-adversarial networks: StyleGAN2-ADA and Pixel2Style2Pixel (Fig.1).

Generative-adversarial networks (GANs) - a machine-learning algorithm without a teacher built on a combination of two neural networks, one of which (generative network) generates samples, and the other one (discriminating network) tries to distinguish correct ("true") samples from the incorrect ones. Generative and discriminating neural networks have the opposite goals - to create samples and to reject samples; an antagonistic game arises between them [1].

StyleGAN (Style Generative Adversarial Network) is an extension of the GAN architecture that maps hidden layers of the network for correlation of points in the latent space with intermediate latent space and uses the intermediate latent space to control the style at each point of the generator model with imputation of noise as a source of variations at each point of the generator model. The resulting model is not only capable of generating impressive photo-realistic high-quality images but also offers style control over the generated image at different levels of detail by changing the vectors of style and noise [2].

According to the StyleGAN producers, it takes 50,000 or more training images to train a high-quality model. Each image must contain a semantic pattern for subsequent reconstruction. The semantic pattern in the given task is a structural element of the relief form or a complex of elements (Fig. 2). Despite the repeatability of these forms within the certain scale range (fractality), it is impossible in this study to provide the necessary number of training images of the required quality using the data on stratigraphic structural surfaces acquired by 3D seismic survey due to relatively small coverage of the project area by 3D seismic. Transfer Learning approach was used to overcome this limitation. Transfer learning is an approach in deep learning, where a new problem is solved by transfer of knowledge from a related problem that has already been studied. One of the reasons for application of this approach is that transfer learning of already trained model (the baseline model) requires a much smaller sampling of new training data from a similar domain. To train the baseline model StyleGAN2-ADA, a digital model of the mountainous surface relief from Shuttle Radar Topography Mission (SRTM) was used, sheet numbers 51-05, 56-02, 56-06. The following image parameters have been determined empirically for the training sample: image size 128x128 pixels, including the area of 12.8x12.8 km; the spectrum from minimum (black) height value to maximum (white) is coded by the grayscale in the window of the selected size. Applying these parameters to the SRTM data, a sampling of 95,338 training images was obtained (Fig. 2).

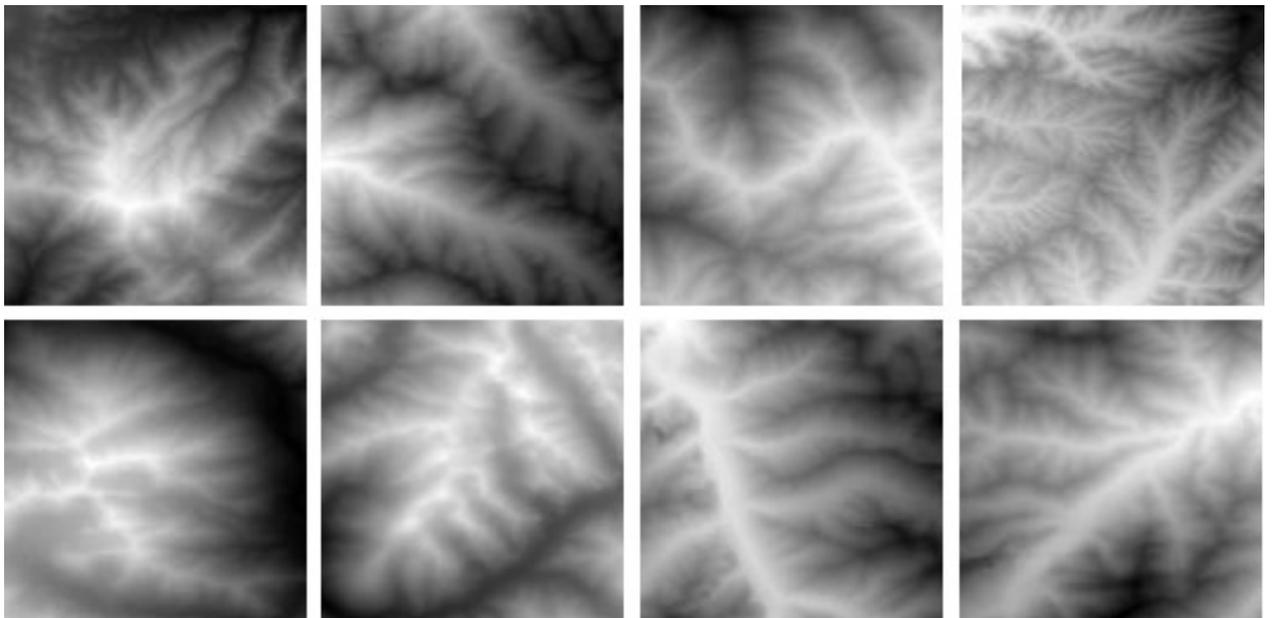

Fig. 2. Example of mountain terrain images from the training sample.

The StyleGAN2-ADA algorithm with the adaptive discriminator extension mechanism was used, which significantly stabilizes learning in modes with limited data [3].
This implementation supports augmentation of training images in process of algorithm training; thus, all available forms of augmentation were used for training, except for colour change. Training was carried out on one graphics processor NVIDIA RTX 2090ti, with default hyperparameter settings and learning rate $3\times10^{-4}$. 105,729 thousand images (kimg) were shown to the best model discriminator during training. Quality of the trained model was determined by the FID metrics. FID is a measure of similarity between two sets of image data. It was shown that it correlates well with the human judgement on the image quality and is most frequently used to assess the quality of samples of generative-adversarial networks [4]. FID is computed by calculating the Frechet distance between two Gaussian distributions suitable for representation of the initial network characteristics. In process of learning, FID reduced from 183 at the beginning of learning to 8.1, by 92,376 kimg; then, after increasing the number of images presented to the discriminator to 105,729 kimg, no change in FID was observed; model training was stopped (Fig. 3); the baseline model for transfer learning (STG2 Mountains Model) was obtained.

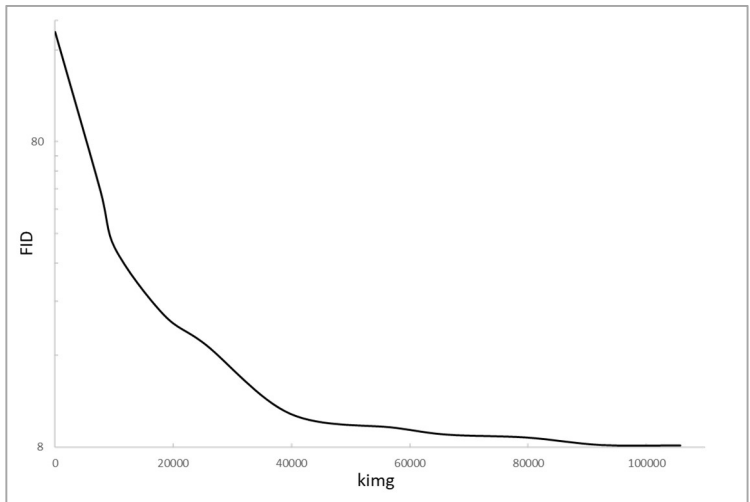

Fig. 3. Change in FID value on a logarithmic scale vs. the number of images displayed to the discriminator when training StyleGAN2-ADA on relief images.

A training sample was created for the transfer learning from the structure maps generated based on the results of 3D seismic survey with the focus on Krasnoleninsk and Frolov petroleum regions of Western Siberia and a target horizon - top of Bazhenov formation. The following parameters have been applied for the images: image size 128x128 pixels covering an area of 12.8x12.8 km, the grayscale encodes the spectrum from minimum (white) value of TVDSS to maximum (black) in the window of the selected size; the image is generated with 3.2 km step horizontally and vertically, thus an increase in sampling is achieved due to overlapping between the areas. The number of acquired images is 154 (Fig. 4). To increase the training sample, a number of transformations were carried out over the images: vertical and horizontal reflections, integer rotations, geometric transformations. A set of 3,155 images has been formed. In case of transfer learning on such a limited dataset, discriminator over-training was observed, resulting in the collapse of adversarial learning. To stabilize learning, images of mountain relief were added to the training sample, with the geological structures similar in shape to the images acquired from 3D seismic. A mixed training sample included 10,000 images.

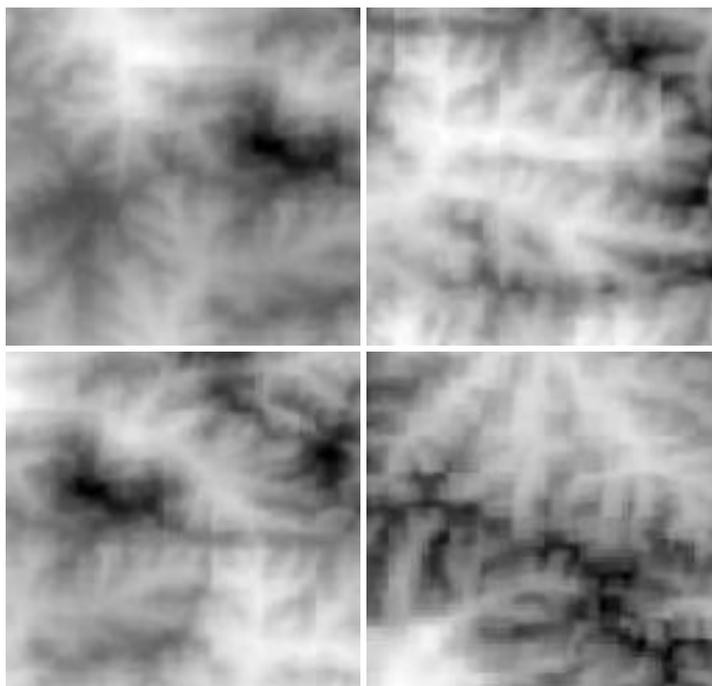

Fig. 4. Example of images based on 3D seismic data from the training sample for transfer learning of StyleGAN2-ADA.

During transfer learning, after the discriminator was shown 21,132 kimg, FID of 12.2 was achieved. Further learning has not changed this metric, and the learning process was stopped at 40,000 kimg. Sizes of the STG2 Transfer Subsurface Model have been obtained for super-resolution model training (Fig. 1).

To implement the super-resolution method, there was used Pixel2Style2Pixel (pSp) algorithm - a StyleGAN2 encoder that uses the multiscale structure of the StyleGAN2 generator to project real images into the hidden space with the possibility of pixel-by-pixel learning using both low- and high-resolution data [5].

Using the STG2 Transfer Subsurface Model, a training dataset of 50,000 images was generated, 80% of which was assigned to the training sample and 20% to the test sample. A trained transfer model (STG2 Transfer Subsurface Model) was used for the encoder, and ResNet-50 model trained on the MOCOv2 dataset was used to calculate the similarity loss function [6].

The choice of the loss function is crucial for the encoder, and the quality of reconstruction directly depends on it. Thus, the pSp encoder is trained using a weighted combination of several functions. A pixel-by-pixel loss function $\mathcal{L}_2$ is used,

$$\mathcal{L}_2(x) = \|x - pSp(x)\|_2,$$

where $x$ means the input image, and $pSp(x) = G(E(x))$ - the output returned to $pSp$, determined by the network of the encoder $E(\cdot)$ and generators network $G(\cdot)$. In order to know the similarity of perception, the loss function $\mathcal{L}_{LPIPS}$ is used, which preserves the image quality in a better way.

$$\mathcal{L}_{LPIPS}(x) = \|F(x) - F(pSp(x))\|_2,$$

where $F(\cdot)$ is the extractor of perceptual features;

$$\mathcal{L}_{MOCO}(x) = 1 - \langle R(x), R(pSp(x)) \rangle,$$

where $R$ - is a neural network of ResNet-50 architecture trained on MOCOv2 data.

Thus, $\mathcal{L}_{MOCO}$ induces the encoder to minimize the cosine similarity between the input features of reconstructed image and its source image. $\mathcal{L}_{MOCO}$ can be applied to any random domain because of the generic nature of the extracted features.

The total loss function is determined as

$$\mathcal{L}(x) = \lambda_1 \mathcal{L}_2(x) + \lambda_2 \mathcal{L}_{LPIPS}(x) + \lambda_3 \mathcal{L}_{MOCO}(x),$$

where, $\lambda_1$, $\lambda_2$, $\lambda_3$ are the constants defining the weights of loss functions.

The model was initiated with the following weights of the loss function $\lambda_1 = 1$, $\lambda_2 = 0.8$, and $\lambda_3 = 0.2$. The pSp algorithm has been modified with respect to the method of down-sampling of images during learning; so, the following sampling filters have been randomly applied: nearest, box, bilinear, hamming, bicubic, lanczos. A down-sampling factor of x16 was used. The other hyperparameters have been kept by default.

Learning was carried out until stabilization of the total loss functions at their minimum values. The resulting loss functions are shown in Table 1.

Table 1. Final values of the loss functions during training the pSp model.

| $\mathcal{L}_2$ | $\mathcal{L}_{LPIPS}$ | $\mathcal{L}_{MOCO}$ | $\mathcal{L}$ |
|---|---|---|---|
| 0.02 | 0.22 | 0.17 | 0.28 |

The loss function values data compared to the results from the paper [6] can be considered acceptable for the goals and objectives of reconstruction by super-resolution method.

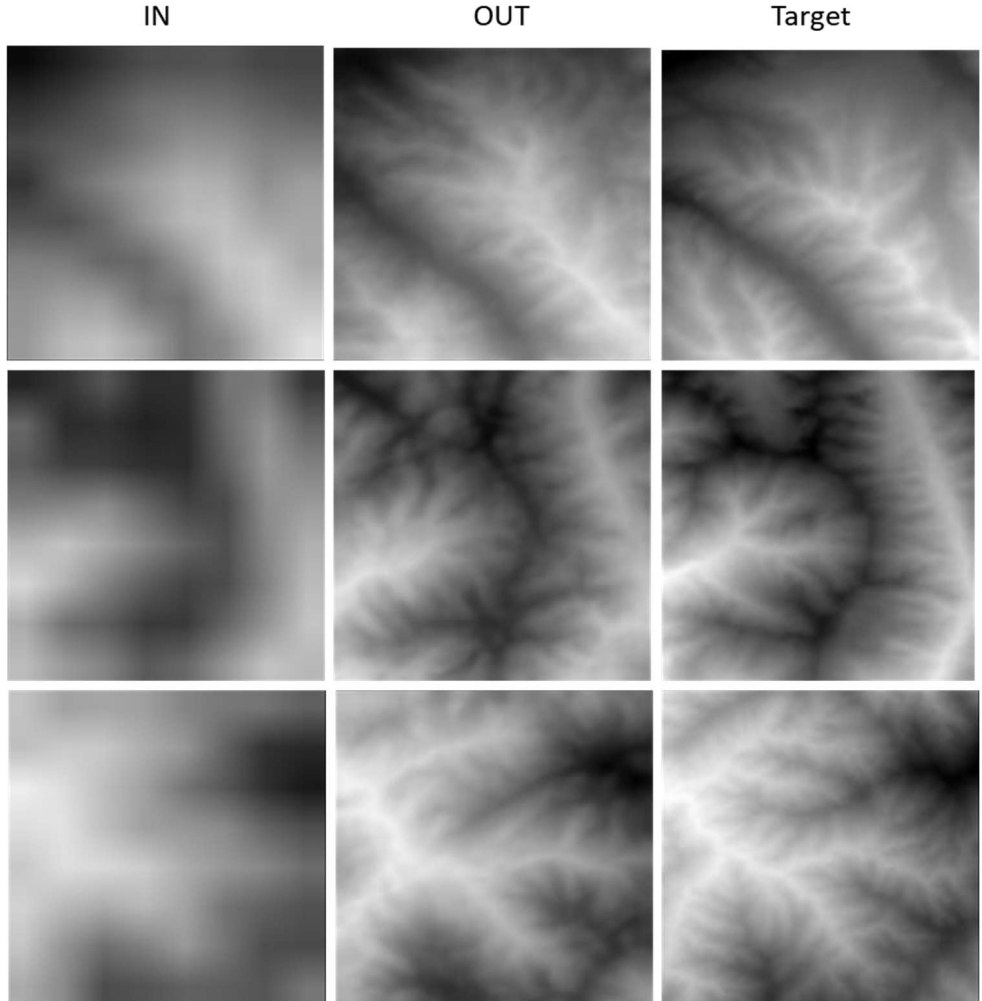

Fig. 5. Example of image reconstruction on the test sample by the pSp algorithm.

Figure 5 shows the model output from the test sample. The "IN" column contains the images sampled from the original image with a factor of x16. The original images are shown in the "Target" column, the "OUT" column shows the results of reconstruction. Visually, there is good quality of reconstruction, given the significant coefficient of sampling of the original image.

## Experiment

Two blocks A and B (Fig. 6), 12.8x12.8 km in size, were selected for the experiment and their data were excluded from the training sample during the transfer learning. Quality assessment of the reconstruction of structural images was carried out on these blocks using the final super-resolution model (pSp) of the target horizon (top of Bazhenov formation). These blocks have a structure geometry characteristic for Krasnoleninsk and Frolov petroleum regions. The distance between them is about 16 km. At different times these areas were successively covered by 2D and 3D seismic data. In 2004, processing and interpretation of 2D seismic data in scope of 278.15 km was carried out for the block A, using the well data from 11 exploration and appraisal wells. 2D seismic data acquisition in scope of 323.48 line km was completed in 2003 in the block B. One well had been drilled in the block by that time. The white lines in Fig. 6 show the 2D seismic lines, and the white dots show the locations of vertical wells. In 2012, a significant part of this area was covered by 3D seismic survey, which included the selected blocks; average fold for the area is 81. By this time, 14 wells had been drilled in Block A and 6 wells in Block B (Fig. 6).

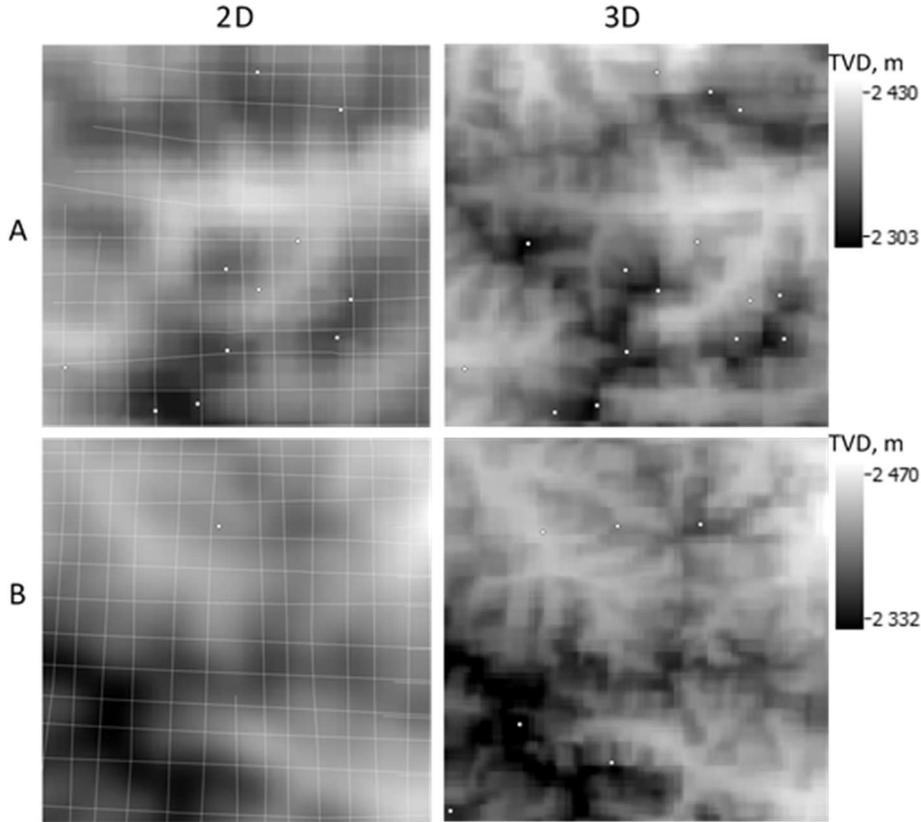

Fig. 6. Depths structure maps based on the results of 2D and 3D seismic data acquisition for the blocks A and B. White lines - 2D seismic lines. White dots - well locations.

TVDSS data for the structural images from 2D seismic were encoded with gray scale from maximum (black colour) to minimum (white colour) and converted to the bitmap image format jpg with 128x128 pixels resolution. The images were input into the trained neural network pSp as low-resolution images with a sampling rate of x16. At the output, the reconstructed image was decoded from the shades of gray to the depths $r$ using the formula

$$r = F(p, L(p, h)), \qquad (1)$$

where $h$ - depth values of the 2D structure map, $p$ — pixel values of the image in the grayscale (0 - 255), $L(\cdot)$ - linear regression coefficient recovery function, $F(\cdot)$ - linear regression function.

Considering the fact that this reconstruction problem does not have a straightforward solution, and comparison of the low-resolution image with the high-resolution image is a "one-to-many" comparison, a multimodal synthesis with a fusion of styles was used to get different variants of credible reconstructions. The multimodal synthesis with style fusion is carried out using the pSp encoder architecture by selecting several style vectors and performing the style mixing at fine-level hidden vectors of the encoded image [5]. 1,000 reconstruction variants were generated for each block.

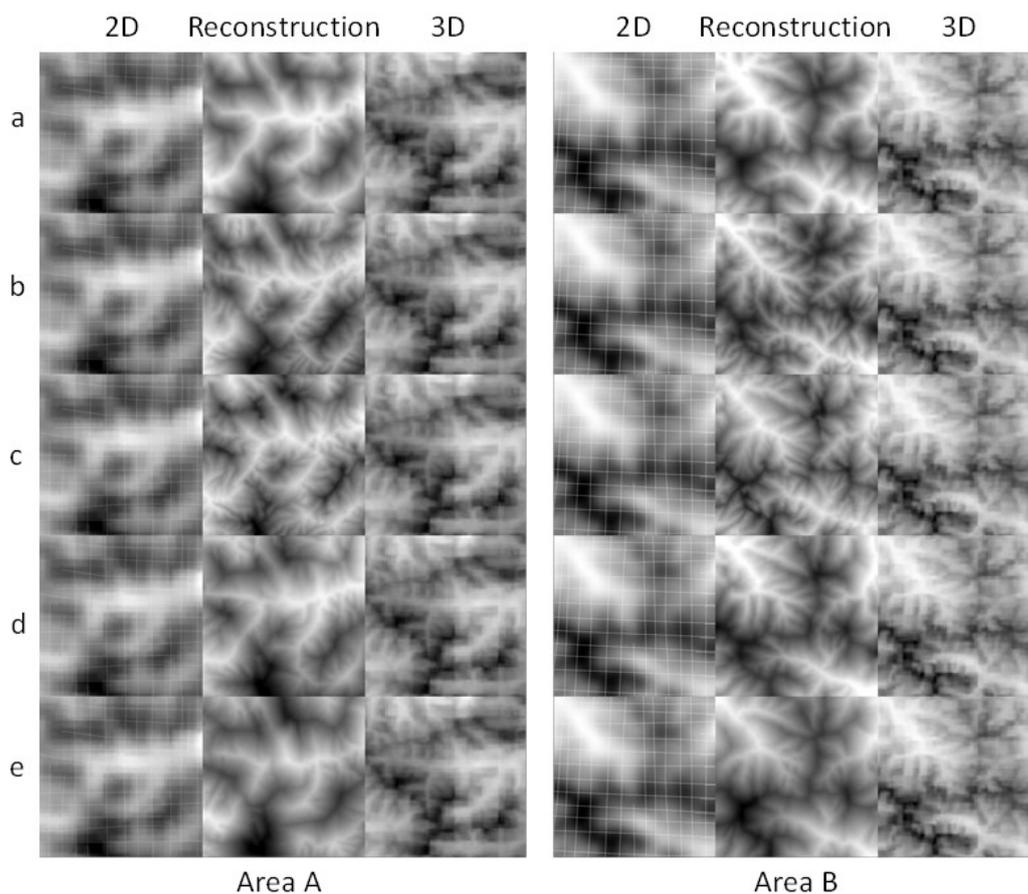

Fig. 7. Examples of reconstruction variations using the multimodal export from the pSp encoder, for Block A - on the left and for Block B - on the right. Column 2D - input image. In the Reconstruction column - multimodal model export in variants: a, b, c, d, e. Column 3D contains the 3D seismic maps. The images are presented in the gray scale unified for each block. White lines are the 2D seismic lines.

In Fig. 7, the Reconstruction column, shows the fluctuation of shapes from variant to variant compared to 2D and 3D seismic maps.

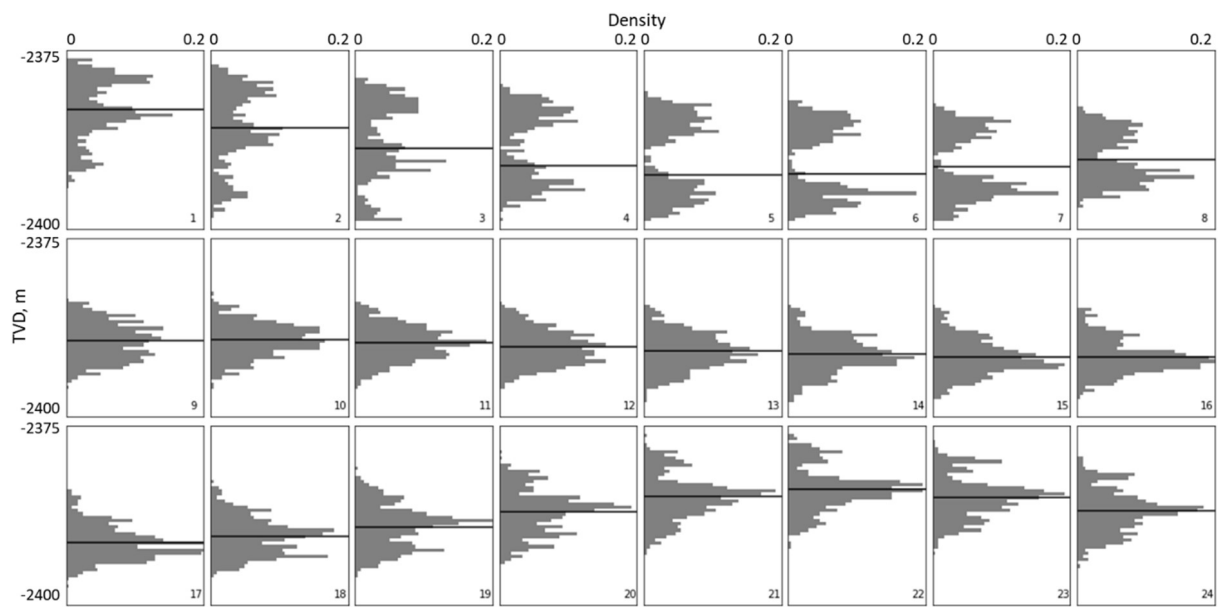

Fig. 8. Densities of the probability distribution of reconstructed depths based on the results of multimodal synthesis in the sequence of 24 points (numerical order from 1 to 24) for the Block B. The black horizontal line on the histograms is the true vertical depth based on 2D seismic in the given point.

Based on the concatenation of the results of multimodal synthesis, a probability space of the project area was formed, where each point is represented by the density of probability distribution of 1,000 equally credible depth values (Fig. 8). Thus, some of the probability distributions have one modality, but some distributions have multiple modality, which is characteristic of the blocks, in which the model has several different versions of reconstruction. Various manipulations with the probability space of the reconstructed images are possible with the use of geostatistics methods. The median depth values (pSp(P50)) have been chosen from the probability space to analyse the quality of reconstructions.

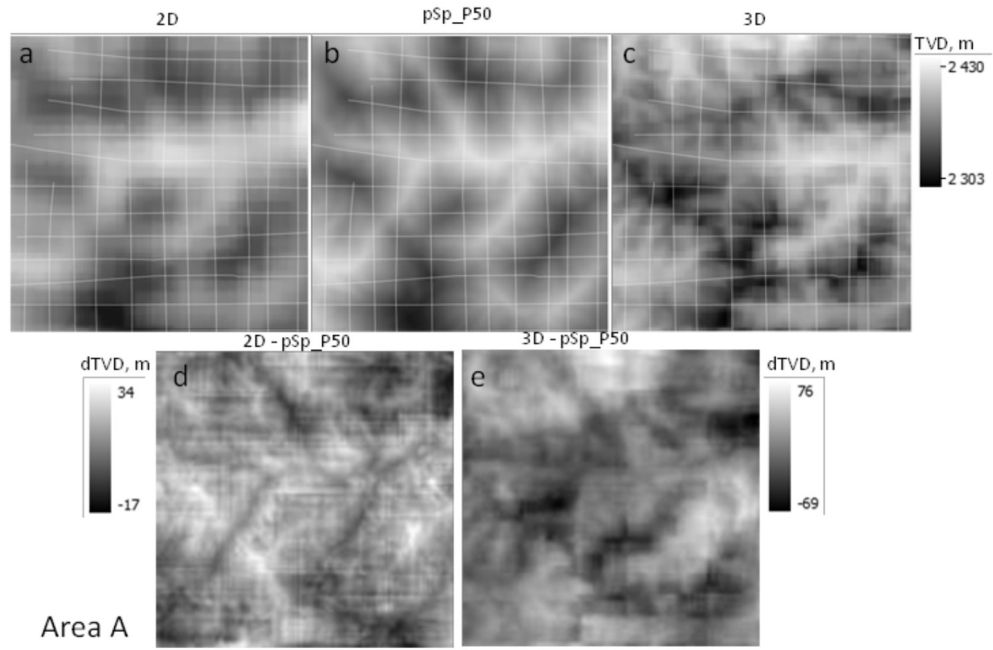

Fig. 9. Block A. TVD maps: a) 2D, b) pSp(P50), c)3D and maps of differences between the original maps and reconstruction: d) 2D-pSp(P50), e) 3D-pSp(P50). White lines are the 2D seismic lines.

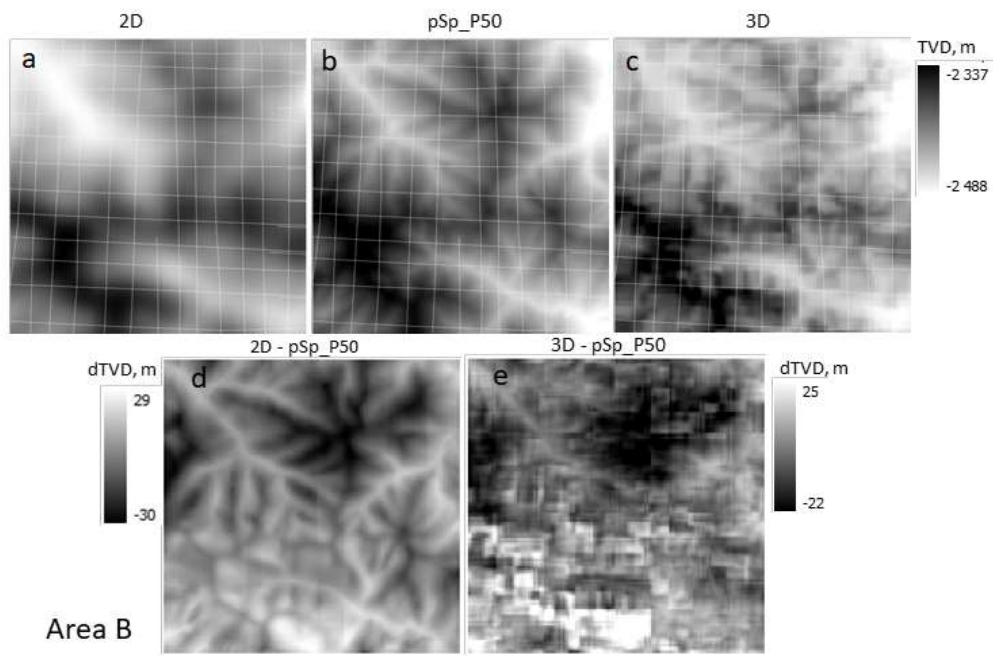

Fig. 10. Block B. TVD maps: a) 2D, b) pSp(P50), c)3D and maps of differences between the original maps and reconstruction: d) 2D-pSp(P50), e) 3D-pSp(P50). White lines – 2D seismic lines.

Fig. 9 and 10 show the TVD maps generated by different methods and the maps showing the difference between the original maps and reconstruction. In general, there is visual similarity between the reconstruction and the original 2D seismic maps (Figs. 9, 10). The model reconstructs very well the detailed structure geometry from the 2D seismic maps. The influence of the surface relief images domain on the structure geometry of reconstructed depth maps is observed. The maps showing the difference between the original maps and reconstruction show both the areas with significant deviations and the areas with relatively small differences.

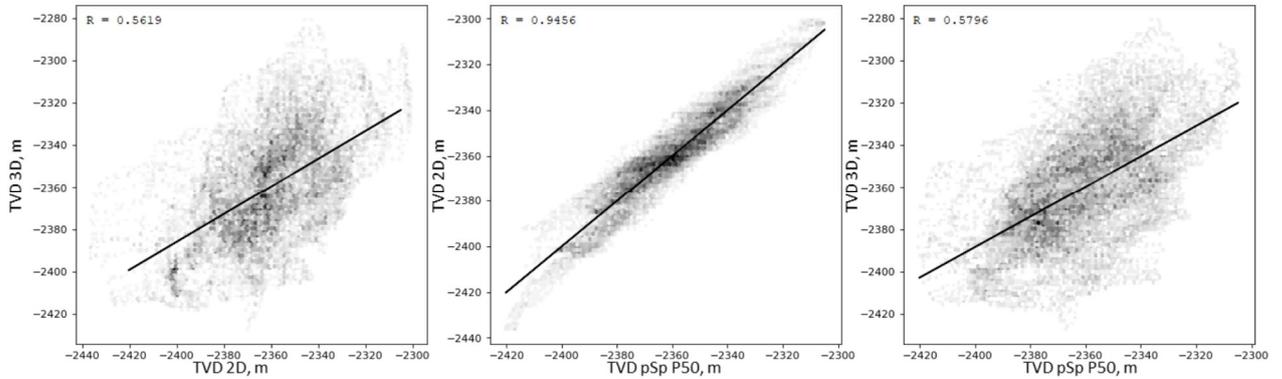

Fig. 11. Cross-plots between the depth values of the maps from left to right: 2D - 3D, pSp(P50) - 2D, pSp(P50) - 3D for the block A

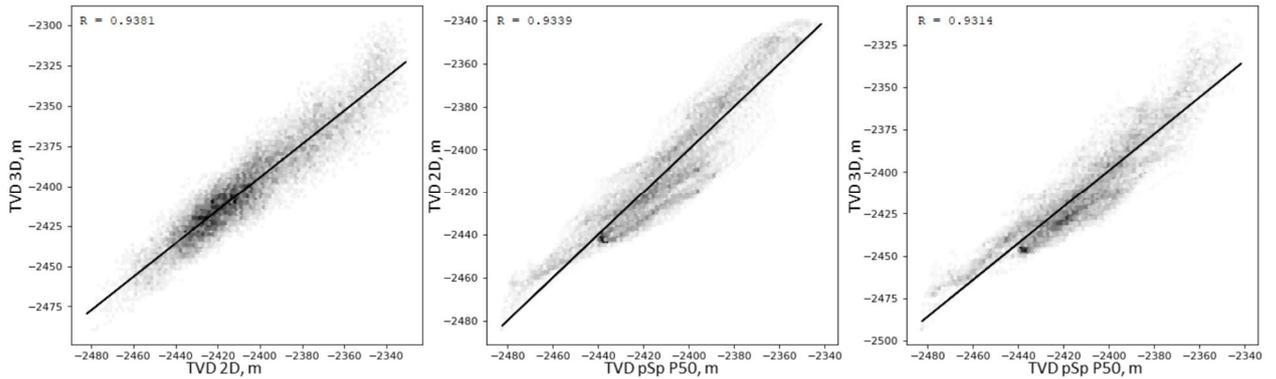

Fig. 12. Cross-plots between the depth values of the maps from left to right: 2D - 3D, pSp(P50) - 2D, pSp(P50) - 3D for the block B.

The cross-plots were built between the pairs of depth values: 2D and 3D, pSp(P50) and 2D, pSp(P50) and 3D; linear regression curves and Pearson coefficient were plotted for each pair (Fig. 11, 12). High correlation coefficient between the original 2D map and the reconstruction (pSp_P50) is due to the method of decoding image colour to depths using the linear regression, according to formula 1. Thus, for the Block A, the Pearson coefficient is 0.9456, for the Block B - 0.9339 (Fig. 11b, Fig. 12b). It is noted that if the correlation between 2D and 3D structure images is low, the reconstructed depths will also poorly correlate with the 3D seismic results (Fig. 11 a, c). At the same time, if the 2D seismic data reproduce the depth trends of the buried relief, the reconstruction algorithm on the basis of semantic images generated during training on high-resolution images picks up the credible geological shapes (Fig. 10 b, c) and preserves the correlation between the depth values (Fig. 12 b, c).

# Conclusions

This paper described a method for reconstruction of detailed-resolution depth structure maps, usually obtained after the 3D seismic surveys, using the data from 2D seismic depth maps. The method uses two algorithms based on the generative-adversarial neural network architecture. The first algorithm StyleGAN2-ADA accumulates in the hidden space of the neural network the semantic images of mountainous terrain forms first, and then with help of transfer learning, in the ideal case – the structure geometry of stratigraphic horizons. The second algorithm, the Pixel2Style2Pixel encoder, using the semantic level of generalization of the first algorithm, learns to reconstruct the original high-resolution images from their degraded copies (super-resolution technology).

There was demonstrated a methodological approach to transferring knowledge on the structural forms of stratigraphic horizon boundaries from the well-studied areas to the underexplored ones. Using the multimodal synthesis of Pixel2Style2Pixel encoder, it is proposed to create a probabilistic depth space, where each point of the project area is represented by the density of probabilistic depth distribution of equally probable reconstructed geological forms of structural images.

Assessment of the reconstruction quality was carried out for two blocks. Using this method, credible detailed depth reconstructions comparable with the quality of 3D seismic maps have been obtained from 2D seismic maps.

It is shown that the quality of reconstruction depends on the quality of the original map generated from 2D seismic data and its correlation with the buried landforms.

This approach can be used as one of the tools for building probabilistic depth maps, along with geostatistical methods, creating ideas about the possible structural and geological features of the area under study.

This method is scalable and after receiving new high-resolution data, with help of the proposed approach in combination with transfer learning, the models can be improved or adapted to specifics of the project area and target stratigraphic horizon.